\documentclass[aps,prl,twocolumn,groupedaddress,showpacs,floatfix]{revtex4}
\usepackage{graphicx}

\begin{document}


\title{Negative thermal expansion of MgB$_{2}$ in the superconducting state and anomalous behavior of the bulk Gr\"uneisen function}

\author{J. J. Neumeier,$^{1}$ T. Tomita,$^{2}$ M. Debessai,$^{2}$ J. S. Schilling,$^{2}$ P. W. Barnes,$^{3}$ D. G. Hinks$^{3}$ and J. D. Jorgensen$^{3}$}
\affiliation{$^{1}$Department of Physics, P.O. Box 173840, Montana State University, Bozeman, Montana 59717-3840, USA}
\affiliation{$^{2}$Department of Physics, Washington University, One Brookings Dr., St. Louis, Missouri 63130-4899, USA}
\affiliation{$^{3}$Materials Science Division, Argonne National Laboratory, Argonne, Illinois 60439, USA}

\date{\today}

\begin{abstract}

The thermal expansion coefficient $\alpha$ of MgB$_2$ is revealed to change from positive to negative on cooling through the superconducting transition temperature $T_c$. The Gr\"uneisen function also becomes negative at $T_c$ followed by a dramatic increase to large positive values at low temperature. The results suggest anomalous coupling between superconducting electrons and low-energy phonons.   

 \end{abstract}

\pacs{65.40.-b, 74.70.Ad, 74.62.Fj}
 
\maketitle

Superconductivity in the binary compound MgB${_2}$ near 39 K is a fascinating development. Over the last few years, scientists have argued that a lattice instability \cite{slusky} and/or anomalous phonon behavior \cite{an} might be responsible for the high transition temperature. Specific attention has focused on the $E_{2g}$ phonon, a bond-stretching phonon within the plane of the hexagonal crystal structure \cite{an,gonc,mialitsin}. In-plane tensile strain, induced by lattice mismatch through thin-film growth, increases the superconducting transition temperature $T_c$ to 41.8 K; this enhancement was attributed to a decrease in the $E_{2g}$ phonon frequency \cite{pogr}. Phonons can be studied with techniques such as Raman spectroscopy \cite{gonc,mialitsin} and heat capacity \cite{wang,bouq,fisher}. Often neglected in the study of phonons is thermal expansion, partly because of the exceptional resolution needed to resolve the transition at $T_c$.

Thermal expansion from powder diffraction measurements \cite{jorg,xue} of MgB$_2$ have revealed an anomalous volume expansion on cooling below $T_c$. However, \textit{high-resolution} thermal expansion measurements (dilatometry) with a relative sensitivity approximately four orders of magnitude better than powder diffraction are required for meaningful thermodynamic analysis. Such measurements of polycrystalline MgB$_{2}$ were reported \cite{lortz}, but discrepancies with the diffraction data, such as the temperature at which the thermal expansion coefficient $\alpha$ changes from positive to negative, are apparent. 

In this Letter, \textit{high-resolution} thermal expansion measurements of polycrystalline MgB$_2$ are presented. The results reveal a change in sign of $\alpha$ at $T_{c}$, with negative thermal expansion below $T_c$; these data agree with diffraction investigations  \cite{jorg,xue}, but offer \textit{exceedingly} greater detail. Analysis of the bulk Gr\"uneisen function reveals anomalous behavior due to dominant low-energy phonon modes. The change in sign of $\alpha$ at \textit{precisely} $T_c$ suggests a connection between these phonon modes and superconductivity.  

MgB$_2$, synthesized with $^{11}$B as described previously \cite{hinks}, was pelletized (diameter = 4.6 mm), placed in a boron nitride crucible and heated to 800 ${^{\circ}}$C for 30 min at 3 GPa using a cubic multi-anvil press. A very thin black layer, impurities from surface reaction with boron nitride, was removed, leaving behind a brilliant gold-colored MgB$_2$ sample with density 2.56 g/cm$^{3}$ (100\% of theoretical density). Heat capacity was measured with a thermal-relaxation technique. $T_c$ versus pressure was determined inductively (0.12 Oe rms field at 1023 Hz) to 0.63 GPa using a helium pressure medium; a manganin sensor at room temperature served as a manometer \cite{tomita}.

Thermal expansion was measured with a novel capacitive dilatometer cell (Fig. 1) constructed from fused quartz \cite{JJN}, which has among the smallest known thermal expansions \cite{quartz}. Calibration is realized by measuring copper \cite{swen} at 300 K and 79.3 K, where the length change of fused quartz, $\Delta$$L$/$L$$_{300 K}$, is zero \cite{quartz}. During measurements,  the sample and cell (in thermal equilibrium) are warmed at 0.2 K/min as capacitance and temperature are recorded at intervals ranging from 0.1 to 0.2 K. The data are corrected for temperature-dependent changes in capacitance of the empty cell, measured in a separate experiment, and the differential expansion of the cell/sample using data for quartz \cite{quartz}. These contributions amount to 2.7 \% and 4.1\%, respectively, of the capacitance change due to the thermal expansion of MgB$_2$ over the temperature range 6 K $<$ $T$ $<$ 300 K; the latter correction would be 63 times larger if a conventional copper dilatometer cell were used to measure MgB$_2$, because of copper's large thermal expansion \cite{swen}. The specimen was measured along two perpendicular axes (4238 $\mu$m and 3665 $\mu$m long) to search for preferred crystallographic orientation effects (none were detected).

\begin{figure}
\includegraphics[scale=0.25]{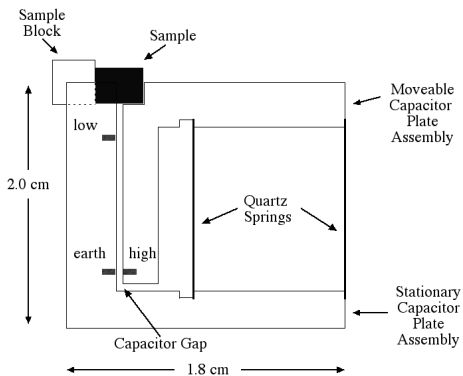}
\vspace{-10pt}
\caption{Composed of two L-shaped pieces connected via 100 $\mu$m thick springs, $\textit{all}$ components of the dilatometer cell are made of fused quartz. Cr/Au (100 \AA/1000 \AA) 1.6 cm$^2$ capacitor plates are perpendicular to the page's plane (electrical contacts shown); a guard ring (at earth) encircles the low plate. A Be-Cu strap (not shown) secures the sample block. Expansion of the sample changes the capacitor gap.  \label{fig1}} 
\vspace{-15pt}
\end{figure}
Linear thermal expansion $\Delta$$L$/$L_{6 K}$ is shown in Fig. 2. These raw data have been corrected for the thermal expansion of quartz as described above with no other processing. The length decreases with temperature, as occurs in most materials, but it expands on cooling below $T_c$ = 38.7 K as illustrated in the inset where the 20 \AA \, scale reveals the \textit{absolute} length change of the 4238 $\mu$m long specimen. The distinct change in $\Delta$$L$/$L_{6 K}$ near $T_c$ is highlighted further in the inset of Fig. 3 where \textit{angstrom-scale} resolution and the continuous (second-order) nature of the phase transition are evident. The measurement was repeated 3 times along the 3665 $\mu$m length and 3 times along the 4238 $\mu$m length with similar results to those shown in Figs. 2 and 3. The overall linear expansion $(L_{300 K}-L_{6 K})$/$L_{6 K}$ = 119.2(4)x10$^{-5}$. 
 
\begin{figure}
\includegraphics[scale=0.36]{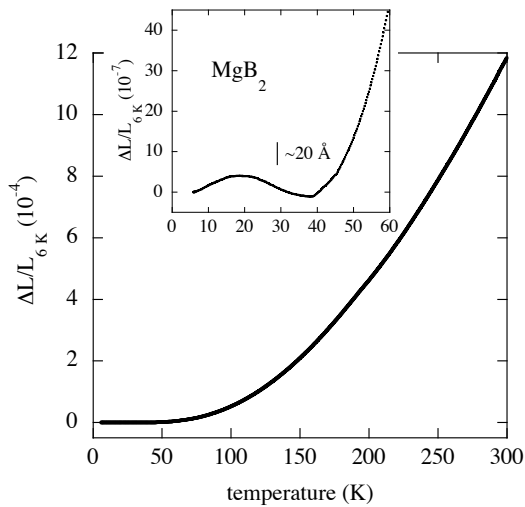}
\vspace{-30pt}
\caption{Linear thermal expansion $\Delta$$L$/$L_{6 K}$ of the 4238 $\mu$m long specimen. The region near $T_c$ is shown in the inset; the 20 \AA\, scale indicates the absolute length change. \label{fig2}} 
\vspace{-15pt}
\end{figure}

The thermal expansion coefficient $\alpha$ = $(1/L_{6 K})$$\partial$$\Delta$$L$/$\partial$$T$ is obtained by fitting the data in Fig. 2 using Chebyshev polynomials and differentiating; overlap between neighboring fit regions assures a smooth derivative and caution is exercised near the sharp feature at $T_c$. This method is typical for taking derivatives when tiny changes between neighboring data points occurs \cite{numer}. The resulting $\alpha$ values are plotted between 0 and 50 K in Fig. 3; all six measurement showed similar behavior. Note that $\alpha$(50 K) is 16.7 times smaller than that of copper \cite{swen}, a consequence of the large Debye temperature of MgB$_2$ \cite{bouq}. Negative thermal expansion sets in at precisely $T_c$ and further inspection reveals a crossover at 18.7 K to $\alpha$ $>$ 0. Furthermore, a broad peak is evident in $\alpha$ at 9.7 K, below which it decreases. According to the third law of thermodynamics, $\alpha$ should approach zero at $T$ = 0; this behavior is illustrated by the (hypothetical) dashed line in Fig. 3.  

The jump in $\alpha$ at $T_c$ (solid lines Fig. 3), $\Delta$$\alpha$ = $\alpha$$_{N}$-$\alpha$$_{S}$ = 6.3(7)x10$^{-8}$ K$^{-1}$,was obtained after averaging the jumps observed in six measurements; $N$ represents the normal state and $S$ the superconducting state. This value agrees with $\Delta$$\alpha$ = 5.8x10$^{-8}$ K$^{-1}$ reported previously \cite{lortz}; in that report, $\alpha$ $<$ 0 below 10 K, but remained above zero for $T$ $>$ 10 K.  These differences may be associated with: (1) the high density of our specimen, since polycrystalline MgB$_{2}$ prepared by the method in Ref.\cite{lortz} is highly porous with low cohesiveness and elastic modulus \cite{harms}, (2) the higher $T_c$ (38.7 K vs. 37.8 K), or (3) the copper thermal expansion cell used in Ref. \cite{lortz} which requires a significantly larger correction to the data than our cell (see above).  Behavior in $\Delta$$L$/$L_{6 K}$ similar to that of Fig. 2 was revealed in diffraction measurements \cite{jorg,xue}, but the present data offer far more detail. 

\begin{figure}
\includegraphics[scale=0.40]{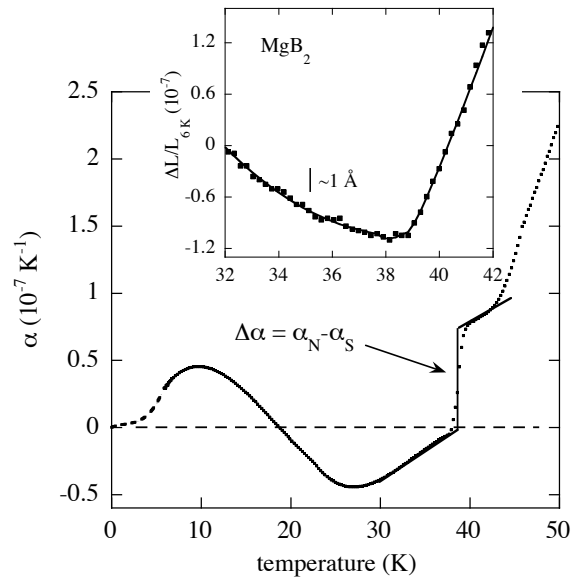}
\vspace{-30pt}
\caption{Inset shows $\Delta$$L$/$L_{6 K}$ near $T_c$ illustrating the change in slope, excellent resolution of the data, and continuous nature of the phase transition.  The solid line in the inset is the fit that, after differentiation, yields the thermal expansion coefficient $\alpha$ (main panel). Solid lines indicate the jump in $\alpha$ at $T_c$. The dashed line below 6 K denotes hypothetical behavior.\label{fig3}}
\vspace{-15pt}
\end{figure}

Heat capacity at constant pressure $C_P$ was measured on the same sample. A background of the form $C_{P}$ = $A_{1}$$T$ + $A_{3}$$T^{3}$ + $A_{5}$$T^{5}$  ($A_{1}$, $A_{3}$, and $A_{5}$ are constants) is subtracted, which was obtained by fitting $C_{P}$(9 tesla) in the range 25 K $<$ $T$ $<$ 50 K; the form of this background is identical to that used previously \cite{bouq}. A clear feature is observed at $T_c$ = 38.7 K (lower inset of Fig. 4). The solid lines denote the step-like behavior expected for a continuous (second-order) phase transition. Using an entropy-conserving construction, the jump $\Delta$$C_{P}$ = -121(1) mJ/mole K is estimated.

The Ehrenfest relation  
\begin{equation}
\frac{dT_{c}}{dP}=3vT_{c}\frac{\Delta \alpha }{\Delta C_{P}},
\end{equation}
provides the connection between the pressure derivative of $T_c$, $\Delta$$\alpha$, and $\Delta$$C_P$ ($v$ is the molar volume). It has yielded $dT_c$/$dP$ in agreement with experiment for other superconductors \cite{meingast,kund}. Using $\Delta$$\alpha$ = 6.3(7)x10$^{-8}$ K$^{-1}$, $\Delta$$C_{P}$ = -121(1) mJ/mole K, and $v$ = 1.74x10$^{-5}$ m$^{3}$/mole \cite{jorg}, $dT_{c}$/$dP$ = -1.05(13) K/GPa is obtained. Variation in $\Delta$$C_P$ values (66 mJ/mole K to 133 mJ/mole K) \cite{fisher} underscores the importance of measuring $\Delta$$C_P$ and $\Delta$$\alpha$ on the same sample to reliably estimate $dT_c$/$dP$. 

\begin{figure}
\includegraphics[scale=0.38]{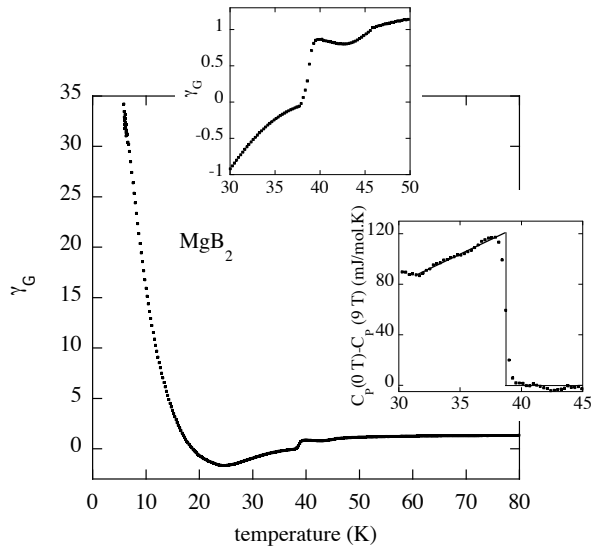}
\vspace{-25pt}
\caption{The main panel and upper inset show the  bulk Gr\"ueneisen function $\gamma$$_G$ versus temperature.  The lower inset shows heat capacity $C_P$ after subtraction of $C_{P}$(9 tesla); the solid line indicates the jump expected for a continuous (second-order) phase transition.\label{fig3}} 
\vspace{-15pt}
\end{figure}

To test the calculation via Eq. (1), $dT_c$/$dP$ was measured on our sample using a \textit{purely hydrostatic method}. Pressure was changed below 55 K and only one transition (P = 0.63 GPa) was measured in solid helium. $T_c$$(P)$ was determined from the midpoint of the superconducting transition in the real part of the ac magnetic susceptibility. The data (Fig. 5) show that $T_c$ decreases at  $dT_{c}$/$dP$ = -1.07(0.03) K/GPa, similar to prior reports \cite{tomita}. This value is in excellent agreement with the value predicted by the Ehrenfest relation. This observation together with the continuous behavior of $\Delta$$L$/$L_{6 K}$ \textit{eliminates the possibility of a significant lattice instability} in MgB$_2$ for 6 K $<$ $T$ $<$ 300 K. 

While negative thermal expansivities are not that unusual \cite{barrera}, the crossover of $\alpha$ from positive to negative, at precisely $T_c$, is rare. Tantalum exhibits a slight negative $\alpha$ \cite{white} below $T_c$ = 4.4 K; what distinguishes this feature in MgB$_2$ is that $\alpha$ is eight times larger above $T_c$, and that it becomes strongly negative below $T_c$. Thus, the electronic transition into the superconducting state has a significant impact on the thermal expansion at and below $T_c$ as revealed in Figs. 2-3.
  
\begin{figure}
\includegraphics[scale=0.36]{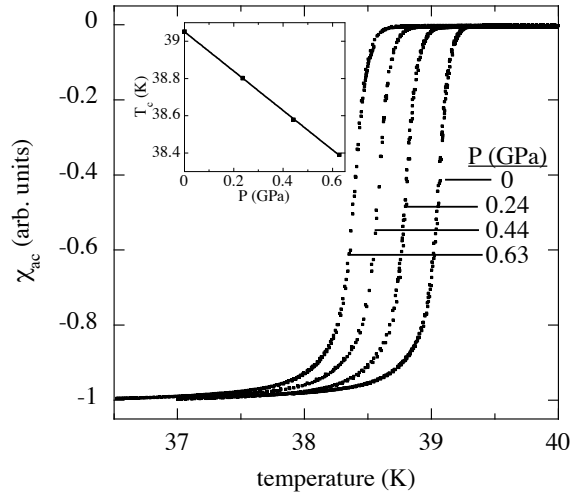}
\vspace{-25pt}
\caption{In the inset $T_c$ versus pressure is plotted.  These data were determined from the midpoint of the transition in the ac magnetic susceptibility $\chi$$_{ac}$, shown in the main panel. \label{fig5}} 
\vspace{-15pt}
\end{figure}

Thermal expansion is a sensitive probe of phonon behavior which is analyzed through the bulk Gr\"uneisen function $\gamma_{G}$ = $\beta$$B$$V$/C$_{V}$ calculated using data \cite{harms} for the bulk modulus $B$ and our measurements of the volume thermal expansion coefficient $\beta$ = 3$\alpha$, $V$ and $C_{P}$ (the heat capacity at constant volume $C_V$ $\approx$ $C_P$ \cite{CP}). It is a dimensionless weighted sum \cite{barrera} of $\gamma$$_i$ values (i.e., $\gamma$$_{G}$ = $\sum_{i}c_i \gamma _i$/$\sum_{i} c_i$) each associated with a phonon mode of vibrational frequency $\omega$$_i$ and $\gamma_i$ = $-dln \omega_i$/$dlnV$. $\gamma$$_{G}$ has a value of 1.30 near 300 K, rises slightly to 1.35 at 90 K followed by a gradual decrease. This decrease continues (see Fig. 4), until $\gamma$$_{G}$  crosses zero at $T_c$, reaches a minimum of -1.65 at 24.5 K, before increasing to 34.2 at 5.8 K. Nonmonotonic behavior of the Gr\"uneisen function is a consequence of strong anharmonic lattice vibrations \cite{abdullaev}, which have been observed in MgB$_2$ \cite{gonc,mialitsin}. The negative as well as the large positive values of $\gamma$$_{G}$ must be associated with anomalous phonon modes that dominate the weighted sum due to their respective thermal populations; electron-phonon interactions will also play a role.

In other layered materials, negative $\alpha$ values are generally associated with transverse acoustic vibrations, which propagate in the layers \cite{abdullaev}, leading to negative a-axis $\gamma_i$ values. As a result, the change in sign of $\alpha$ and $\gamma$$_{G}$ at $T_c$, coincident with hardening of the E$_{2g}$ phonon \cite{mialitsin}, is likely connected to planar electron-phonon interactions associated with the $\sigma$-band (in-plane) superconducting gap \cite{choi, souma}. 

At low temperature, differences in the intralayer bonding (along the c-axis) may lead to low-energy phonon excitations and large c-axis $\gamma_i$ values with high relative thermal population that cause the upturn of $\gamma_{G}$ below 20 K; such behavior occurs in other layered materials, albeit with significantly smaller values of  $\gamma_{G}$ \cite{abdullaev}. In MgB$_2$, the spectacular increase of $\gamma$$_{G}$ coincides with a reduced $C_P$ below $\approx$ 11 K, attributed to a second superconducting gap \cite{wang, bouq}. Thus, $C_P$ is diminished because of electronic effects due to the second gap, $\alpha$ remains large, and their ratio leads to anomalously large $\gamma$$_{G}$ values. These combined considerations suggest that the increase in $\gamma$$_{G}$ at low temperature may result from out-of-plane electron-phonon interactions, associated with the weaker $\pi$-band superconducting gap \cite{choi, souma}, which have a strong impact on the c-axis $\gamma_i$ values.  

The results presented here compliment spectroscopic phonon probes, some of which may be surface sensitive, by illustrating unusual phonon behavior in two fundamental \textit{bulk} thermodynamic quantities, the Gr\"uneisen function and the thermal expansion coefficient. Thermal excitations of electrons across the two superconducting gaps, and their coupling to low-energy lattice vibrations, play a role in the anomalous behavior of $\gamma_G$ and $\alpha$. 

\begin{acknowledgments}
Thanks to J. Macaluso (Research Experiences for Undergraduates student) for constructing the dilatometer cell, R. Bollinger and H. Terashita for assistance, W. E. Pickett, A. Serquis and V. F. Nesterenko for comments. This material is based upon work supported by the National Science Foundation (Grant No. DMR 0504769, DMR 0301166, DMR 0244058 (MSU) and DMR 0404505 (Washington University)). Work at ANL is supported by the Department of Energy under Contract No. W-31-109-ENG-38.    
\end{acknowledgments}

\end{document}